# A temperate rocky super-Earth transiting a nearby cool star

Jason A. Dittmann[1], Jonathan M. Irwin[1], David Charbonneau[1], Xavier Bonfils[2,3], Nicola Astudillo-Defru[4],Raphaëlle D. Haywood[1], Zachory K. Berta-Thompson[5], Elisabeth R. Newton[6], Joseph E. Rodriguez[1], Jennifer G. Winters[1], Thiam-Guan Tan[7], Jose-Manuel Almenara[2,3,4], François Bouchy[8], Xavier Delfosse[2,3], Thierry Forveille[2,3], Christophe Lovis[4], Felipe Murgas[2,3,9], Francesco Pepe[4], Nuno C. Santos[10,11], Stephane Udry[4], Anaël Wünsche[2,3], Gilbert A. Esquerdo[1],David W. Latham[1] & Courtney D. Dressing[12]

**M dwarf stars, which have masses less than 60 per cent that of the Sun, make up 75 per cent of the population of the stars in the Galaxy[1]. The atmospheres of orbiting Earth-sized planets are observationally accessible via transmission spectroscopy when the planets pass in front of these stars[2,3]. Statistical results suggest that the nearest transiting Earth-sized planet in the liquid-water, habitable zone of an M dwarf star is probably around 10.5 parsecs away[4]. A temperate planet has been discovered orbiting Proxima Centauri, the closest M dwarf[5], but it probably does not transit and its true mass is unknown. Seven Earth-sized planets transit the very low-mass star TRAPPIST-1, which is 12 parsecs away[6,7], but their masses and, particularly, their densities are poorly constrained. Here we report observations of LHS 1140b, a planet with a radius of 1.4 Earth radii transiting a small, cool star (LHS 1140) 12 parsecs away. We measure the mass of the planet to be 6.6 times that of Earth, consistent with a rocky bulk composition. LHS 1140b receives an insolation of 0.46 times that of Earth, placing it within the liquid-water, habitable zone[8]. With 90 per cent confidence, we place an upper limit on the orbital eccentricity of 0.29. The circular orbit is unlikely to be the result of tides and therefore was probably present at formation. Given its large surface gravity and cool insolation, the planet may have retained its atmosphere despite the greater luminosity (compared to the present-day) of its host star in its youth[9,10]. Because LHS 1140 is nearby, telescopes currently under construction might be able to search for specific atmospheric gases in the future[2,3].**

[1]Harvard Smithsonian Center for Astrophysics, 60 Garden Street, Cambridge, Massachusetts 02138, USA. [2]CNRS (Centre National de la Recherche Scientifique), IPAG (Institut de Planétologieet d'Astrophysique de Grenoble), F-38000 Grenoble, France. [3]Université Grenoble Alpes, IPAG, F-38000 Grenoble, France. [4]Observatoire de Genève, Université de Genève, 51 chemin des Maillettes, 1290 Versoix, Switzerland. [5]University of Colorado, 391 UCB, 2000 Colorado Avenue, Boulder, Colorado 80305, USA. [6]Massachusetts Institute of Technology, 77 Massachusetts Avenue, Cambridge, Massachusetts 02138, USA. [7]Perth Exoplanet Survey Telescope, Perth, Western Australia, Australia. [8]Aix Marseille Université, CNRS, LAM (Laboratoire d'Astrophysique de Marseille) UMR 7326, 13388 Marseille, France. [9]Instituto de Astrofísica de Canarias (IAC), E-38205 La Laguna, Tenerife, Spain. [10]Instituto de Astrofísica e Ciências do Espaço, Universidade do Porto, CAUP (Centro de Astrofísica da Universidade do Porto), Rua das Estrelas, 4150-762 Porto, Portugal. [11]Departamento de Física e Astronomia, Faculdade de Ciências, Universidade do Porto, Rua do Campo, Alegre, 4169-007 Porto, Portugal. [12]Division of Geological and Planetary Sciences, California Institute of Technology, Pasadena, California 91125, USA.

MEarth[11,12] consists of two arrays of eight 40-cm-aperture telescopes, one in the Northern Hemisphere at the Fred Lawrence Whipple Observatory (FLWO) in Arizona, USA, and the other in the Southern Hemisphere at Cerro Tololo Inter-American Observatory, Chile. This survey monitors small stars (less than 33% the size of the Sun) that are estimated to lie within 100 light years of the Sun for transiting extrasolar planets. Since January 2014, these telescopes have gathered data nearly every clear night, monitoring the brightnesses of these stars for signs of slight dimming, which would be indicative of a planet transiting in front of the star. MEarth-South monitors these stars at a cadence of approximately 30 min, but is capable of performing high-cadence observations in real time if a possible planetary transit is detected to be in progress[13]. We used the MEarth-South telescope array to monitor the brightness of the star LHS 1140, beginning in 2014.

The distance to LHS 1140 has been measured through trigonometric parallax to be 12.47 ± 0.42 parsecs (ref. 14). Combined with the 2MASS $K_s$ magnitude[15] and empirically determined stellar relationships[16,17], we estimate the stellar mass to be 14.6% that of the Sun and the stellar radius to be 18.6% that of the Sun. We estimate the metal content of the star to be approximately half that of the Sun ([Fe/H] = –0.24 ± 0.10; $1\sigma$ error), and we measure the rotational period of the star to be 131 days from our long-term photometric monitoring (see Methods).

On 15 September 2014 UT, MEarth-South identified a potential transit in progress around LHS 1140, and automatically commenced high-cadence follow-up observations (see Extended Data Fig. 1). Using a machine-learning approach (see Methods), we selected this star for further follow-up observations.

We gathered two high-resolution (resolution $R$ = 44,000) reconnaissance spectra with the Tillinghast Reflector Echelle Spectrograph (TRES) on the 1.5-m Tillinghast reflector located at the Fred Lawrence Whipple Observatory (FLWO) on Mt Hopkins, Arizona, USA. From these spectra, we ruled out contamination from additional stars and large systemic accelerations, and concluded that this system was probably not a stellar binary or false positive (see Methods). We subsequently obtained 144 precise radial velocity measurements with the High Accuracy Radial Velocity Planet Searcher (HARPS) spectrograph[18] from 23 November 2015 to 13 December 2016 UT.

On 19 June 2016 UT, MEarth-South detected an additional transit of LHS 1140b through the MEarth trigger; when combined with the radial velocities and our initial trigger, we identified three potential orbital periods. On the basis of one of these, the 24.738-day period, we back-predicted a third, low-signal-to-noise transit from 23 December 2014 UT (see Extended Data Fig. 1). With this ephemeris, we predicted a transit on 01 September 2016 UT, the egress of which was observed by the Perth Exoplanet Survey Telescope (PEST; see Methods). On 25 September 2016 and 20 October 2016 UT we obtained two complete transit observations with four of the eight MEarth-South telescopes.

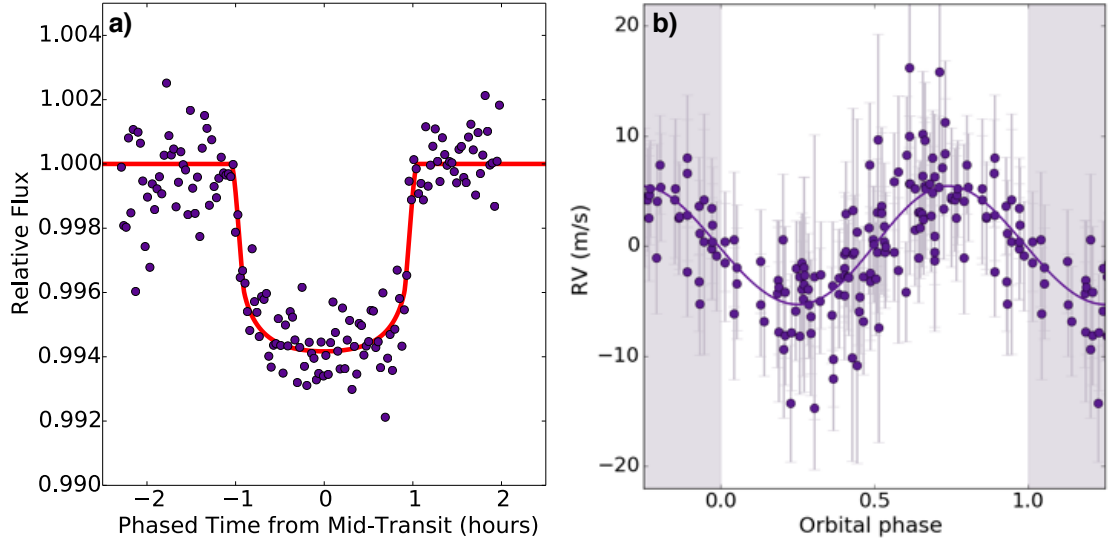

*Figure 1 | Photometric transit and radial-velocity measurements of LHS 1140b. **a**, Phase-folded transit observations from all transits (purple), with our transit model over-plotted as a red line. These data were binned in 160 3-min bins. Here we have corrected each individual light curve from each telescope with a zero-point offset, and with a linear correction for the air mass of the observation. For both full-transit observations we also apply a correction that is linear in time so that the flux level of LHS 1140 is equal before and after the transit. **b**, 144 measurements of the line-of-sight (radial) velocity of LHS 1140 taken with the HARPS spectrograph (purple points, duplicate observations are shown in the shaded regions, error bars are $1\sigma$). A value of zero corresponds to a radial velocity equal to that of the host star. We have removed variability due to stellar activity and plot only the radial-velocity perturbations induced by the planet, and have phase-folded the radial velocities to the orbital period. Our best-fitting Keplerian orbit is shown as the solid purple line.*

We initially fitted our photometric transit and radial-velocity measurements simultaneously (see Methods). This fit serves as input to a comprehensive radial-velocity analysis that takes into account not only the reflex motion of the planet, but also the intrinsic variations of the host star via Gaussian process regression[19,20] (see Methods). We find that LHS 1140b has a mass 6.65 ± 1.82 times that of Earth and a radius 1.43 ± 0.10 times that of Earth, and orbits around LHS 1140 with a period of 24.73712 ± 0.00025 days (all error bars are $1\sigma$) and an eccentricity that is constrained to be less than 0.29 (at 90% confidence; see Methods). In Fig. 1, we show the phased and binned transit light curve for LHS 1140b and the phased radial-velocity curve; in Extended Data Fig. 1 we show each individual transit. Our phased radial-velocity curve shows the radial velocity from the influence of LHS 1140b only; the stellar contribution has been removed. In Table 1, we show the system parameters for LHS 1140 and LHS 1140b, with 68% confidence intervals on these parameters.

A simple structural model surrounded by a magnesium silicate mantle can explain the observed mass and radius (see Fig. 2). Although our best fitting values imply a much higher iron core-mass fraction than that of Earth (0.7 compared to 0.3), our uncertainties on the mass and the radius can only rule out Earth- like compositions at $2\sigma$ confidence. We conclude that LHS 1140b is a rocky planet without a substantial gas envelope.

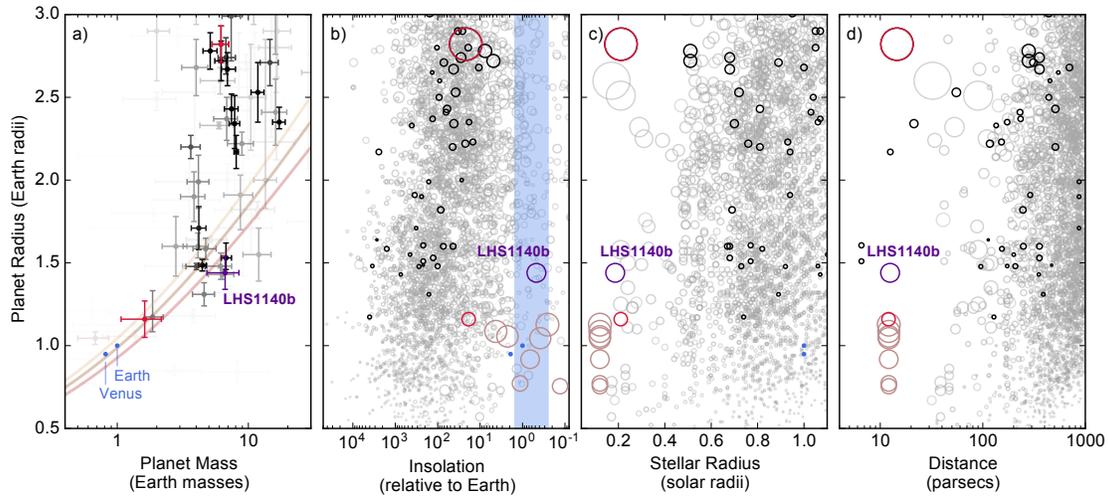

*Figure 2 | Masses, radii, distances, insolation and stellar size of known transiting planets.* **a**, The mass and radius of LHS 1140b indicate a terrestrial composition. Other planets with measured masses and radii are shown, with darker points indicating smaller density uncertainties. The red points correspond to GJ 1214b (top) and GJ 1132b (below LHS 1140b). Error bars, 1σ. Mass–radius curves for two-layer rocky planets with 0%, 25% and 50% of their mass in iron cores are shown as solid lines. **b–d**, Planetary radius versus insolation, stellar radius and distance, respectively. Planets with dynamical mass determinations are shown in black; those without are shown in grey. The red data correspond to GJ 1214b and GJ 1132b, as in **a**, and the darker red circles to the TRAPPIST-1 planets; these are the nearby planets around small stars that are most accessible to characterization by the James Webb Space Telescope (JWST), as indicated in **c** and **d**. The shaded region in **b** is the M-dwarf habitable zone[8]. We note that this habitable zone is only appropriate for planets orbiting M dwarfs, and most of the planets in this diagram orbit much larger stars. The area of each circle is proportional to the transit depth and hence observational accessibility. LHS 1140b has a lower insolation than Earth, and orbits a small star 12 parsecs from the Sun, making it a temperate, rocky planet that may be accessible to atmospheric characterization.

We searched for additional planets both in the MEarth-South light curve and as periodic signals in the HARPS residuals. We did not find any compelling signals in these data. After subtracting the radial-velocity signal due to stellar rotation and LHS 1140b, the Lomb–Scargle periodogram of the residuals shows a series of broad peaks at periods greater than 60 days, possibly associated with stellar activity. At shorter periods, the highest peaks are the result of the window function of our radial-velocity observations (see Methods). We do not find any notable periodic signals or additional triggers that would be suggestive of another transiting planet, although planets smaller than LHS 1140b could elude detection.

Compact, coplanar, multi-planet systems are common around M dwarfs[22,23], and all coplanar objects at periods of less than the orbital period of LHS 1140b would also transit, although their size may be too small to have been detected. The surface density of the protoplanetary disk in which LHS 1140b formed may dictate the properties of any additional planets in this system. A high-surface-density disk may shorten the timescale for planet formation to be before the dissipation of the gaseous disk component, allowing otherwise terrestrial planets to accrete large envelopes of hydrogen and helium[24]. LHS

1140b does not have such a gaseous component, so it might have formed from a low-surface-density disk, consistent with the low metallicity of the star, although in this scenario we would expect additional rocky worlds. Further radial-velocity and photometric monitoring of LHS 1140 is warranted.

A recent study[8] found that a planet orbiting an M dwarf could have surface temperatures that allow liquid water if it receives between 0.2 and 0.8 times the insolation that Earth receives from the Sun. LHS 1140b currently receives 0.46 times Earth's insolation, and we estimate its age to exceed 5 Gyr (see Methods). In its youth, LHS 1140 was more luminous, and a larger fraction of its spectrum was released at ultraviolet wavelengths. During this period, the atmosphere of LHS 1140b was therefore subjected to increased irradiance and greater levels of ionizing radiation, and LHS 1140b probably did not enter the liquid-water, habitable zone until approximately 40 Myr after the formation of the star[9]. This amount of time may have been sufficient for the atmosphere to have experienced a runaway greenhouse, with water being dissociated in the upper atmosphere and the hydrogen permanently lost to atmospheric escape[9]. If so, then the planet's atmosphere would be dominated by abiotic $O_2$, $N_2$ and $CO_2$. However, recent work has suggested that super-Earths can have an extended magma-ocean phase[10], in which case the timescale over which LHS 1140b outgassed its secondary atmosphere may have exceeded the time for the star to reach its current luminosity. In this scenario, volatiles such as $H_2O$ would have remained in the mantle of the planet until after the host star dimmed and its fractional ultraviolet emission decreased. Inferences of the history of the atmosphere would be strengthened with better observational constraints on the emission of M dwarfs at young ages, and with more detailed models of the initial composition of the atmosphere from outgassing and the delivery of volatiles through late-stage cometary impacts. Observations of the current ultraviolet emission of LHS 1140 by the Hubble Space Telescope will be able to be used to assess the current high-energy flux that is infringing upon LHS 1140b, and will be helpful in determining the current habitability of LHS 1140b and constraining any ongoing atmospheric escape from the planet.

## Table 1: System Parameters for LHS 1140

| Parameter | Value (median and 1-σ error) |
| --- | --- |
| **Stellar Parameters** | |
| **RA (J2000)** | 00:44:59.3 |
| **DEC (J2000)** | -15:16:18 |
| **Proper Motion** | 665.9 ± 1.0 mas yr$^{-1}$ |
| **Position Angle** | 153.3 ± 0.16 deg |
| **Photometry** | $V = 14.18 ± 0.03$, $R = 12.88 ± 0.02$, $I_c = 11.19 ± 0.02$, $J = 9.612 ± 0.023$, $H = 9.092 ± 0.026$, $K_s = 8.821 ± 0.024$, $W1 = 8.606 ± 0.022$, $W2 = 8.391 ± 0.019$, $W3 = 8.235 ± 0.022$, $W4 = 8.133 ± 0.265$ |
| **Distance to star, $D_*$** | 12.47 ± 0.42 parsecs |
| **Mass of star, $M_*$** | 0.146 ± 0.019 $M_\odot$ |
| **Radius of star, $R_*$** | 0.186 ± 0.013 $R_\odot$ |
| **Luminosity of star, $L_*$** | 0.002981 ± 0.00021 $L_\odot$ |
| **Effective temperature, $T_{eff}$** | 3131 ± 100 K |
| **Metallicity, [Fe/H]** | -0.24 ± 0.10 (± 0.1, systematic) |
| **Age of star, $\tau_*$** | > 5 Gyr |
| **$P_{rotation}$** | 131 days |
| **Systemic velocity, $\gamma_*$** | -13.23 ± 0.60 km s$^{-1}$ |
| | |
| **Transit and Radial Velocity Parameters** | |
| **Planet-to-star radius ratio, $R_p / R_*$** | 0.0708 ± 0.0013 |
| **Inclination (degrees), $i$** | 89.912 ± 0.071 |
| **RV semi-amplitude** | 5.34 ± 1.1 m s$^{-1}$ |
| **Eccentricity, $e$** | < 0.29 (90% confidence) |
| **Period, $P$** | 24.73712 ± 0.00025 days |
| **Time of mid-transit, $T_0$** | 2456915.6997 ± 0.0054 HJD |
| **Scaled orbital distance, $a/R_*$** | 101.0 ± 6.0 |
| | |
| **Derived Parameters** | |
| **Mass of planet, $M_p$** | 6.65 ± 1.82 $M_\oplus$ |
| **Radius of planet, $R_p$** | 1.43 ± 0.10 $R_\oplus$ |
| **Density of planet, $\rho_p$** | 12.5 ± 3.4 g cm$^{-3}$ |
| **Escape velocity, $v_{esc}$** | 24.0 ± 2.7 km s$^{-1}$ |
| **Surface gravity of planet, $g_p$** | 31.8 ± 7.7 m s$^{-2}$ |
| **Equilibrium Temperature, $T_{eq}$ (0 albedo)** | 230 ± 20 K |
| **Semi-major axis** | 0.0875 ± 0.0041 AU |

RA, right ascension; dec., declination; $M_\odot$, $R_\odot$ and $L_\odot$ are the mass, radius and luminosity, respectively, of the Sun; $M_\oplus$ and $R_\oplus$ are the mass and radius, respectively, of Earth; HJD, heliocentric Julian date.

**Supplementary Information** is available in the online version of the paper.

**Acknowledgements** We thank the staff at the Cerro Tololo Inter-American Observatory for assistance in the construction and operation of MEarth-South. The MEarth team acknowledges funding from the David and Lucille Packard Fellowship for Science and Engineering (awarded to D.C.). This material is based on work supported by the National Science Foundation under grants AST-0807690, AST-1109468, AST-1004488 (Alan T. Waterman Award) and AST-1616624. This publication was made possible through the support of a grant from the John Templeton Foundation and NASA XRP Program #NNX15AC90G. The opinions expressed in this publication are those of the authors and do not necessarily reflect the views of the John Templeton Foundation. HARPS observations were made with European Southern Observatory (ESO) telescopes. This work was performed in part under contract with the Jet Propulsion Laboratory (JPL) funded by NASA through the Sagan Fellowship Program executed by the NASA Exoplanet Science Institute. E.R.N. is supported by an NSF Astronomy and Astrophysics Postdoctoral Fellowship under award AST-1602597. N.C.S. acknowledges support from Fundação para a Ciência e a Tecnologia (FCT) through national funds and by FEDER through COMPETE2020 by grants UID/FIS/04434/2013&POCI-01-0145-FEDER-007672 and PTDC/FIS-AST/1526/2014&POCI-01-0145-FEDER-016886. N.C.S. was also supported by FCT through InvestigadorFCT contract reference IF/00169/2012/CP0150/CT0002. X.B., X.D. and T.F. acknowledge the support of the INSU/PNP (Programme national de planétologie) and INSU/PNPS (Programme national de physique stellaire). X.B., J.-M.A. and A.W. acknowledge funding from the European Research Council under ERC Grant Agreement no. 337591-ExTrA. We thank A. Vanderburg for backseat MCMCing. This publication makes use of data products from the Two Micron All Sky Survey (2MASS), which is a joint project of the University of Massachusetts and the Infrared Processing and Analysis Center/California Institute of Technology, funded by NASA and the National Science Foundation. This publication makes use of data products from the Wide-field Infrared Survey Explorer, which is a joint project of the University of California, Los Angeles, and the JPL/California Institute of Technology, funded by NASA. This research has made extensive use of the NASA Astrophysics Data System (ADS), and the SIMBAD database, operated at CDS, Strasbourg, France.

**Author Contributions** The MEarth team (J.A.D., D.C., J.M.I., Z.K.B.-T., E.R.N., J.G.W. and J.E.R.) discovered the planet, organized the follow-up observations, and led the analysis and interpretation. J.A.D. analysed the light curve and the radial-velocity data and wrote the manuscript. J.M.I. designed and installed, and maintains and operates the MEarth-South telescope array, and contributed to the analysis and interpretation. D.C. leads the MEarth project, and assisted in analysis and writing the manuscript. E.R.N. determined the rotational period of the star. R.D.H. conducted the Gaussian process analysis of the radial velocities. J.E.R. and T.-G.T. organized the follow-up effort in Perth. The HARPS team (X.B., N.A.-D., J.-M.A., F.B., X.D., T.F., C.L., F.M., F.P., N.C.S., S.U. and A.W.) obtained spectra for Doppler velocimetry, with N.A.-D. and X.B. leading the analysis of those data. G.A.E. and D.W.L. obtained the reconnaissance spectrum with TRES at FLWO. C.D.D. obtained the infrared spectrum with IRTF/SpeX and determined the stellar metallicity.

**Author Information** Reprints and permissions information is available at www.nature.com/reprints. The authors declare no competing financial interests. Readers are welcome to comment on the online version of the paper. Publisher's note: Springer Nature remains neutral with regard to jurisdictional claims in published maps and institutional affiliations. Correspondence and requests for materials should be addressed to J.A.D. (Jason.Dittmann@gmail.com).

**Reviewer Information** *Nature* thanks A. Hatzes and the other anonymous reviewer(s) for their contribution to the peer review of this work.

# METHODS

**Mass of the star.** Dynamical mass measurements of M-dwarf visual binaries, combined with accurate distance and magnitude measurements, have shown that there is a precise relationship between the near-infrared absolute magnitude of an M dwarf and its mass[16,25]. We used this relation[16] to calculate the mass of LHS 1140 from its $K_s$ magnitude, because the relationship in $K_s$ is the most precise. We adopt a mass of $M_* = (0.146 \pm 0.019) M_\odot$, with the uncertainty of the mass primarily determined by the scatter in the mass–luminosity relation.

**Radius of the star.** From this mass, we use a mass–radius relation calibrated with long-baseline optical interferometry of single stars[17] to calculate the stellar radius: $R_* = (0.186 \pm 0.013) R_\odot$. The Dartmouth Stellar Evolution Database[26] offers a suite of stellar models across stellar mass, age and composition. Interpolating those models for a star of LHS 1140's mass with [Fe/H] = 0 and [α/Fe] = 0 yields a stellar radius of $(0.166 \pm 0.11) R_\odot$. Similarly, a model track with [Fe/H] = –0.5 and [α/Fe] = 0 yields a stellar radius of $(0.163 \pm 0.010) R_\odot$, consistent with the empirical relation that we have adopted. Additional stellar models[27] at an age of 5.0 Gyr estimate a stellar radius of $(0.166 \pm 0.09) R_\odot$, but under-predict the absolute $K_s$-band magnitude by 0.15 magnitudes. Inflating the model radius to correct the absolute $K_s$ magnitude to the observed value increases the stellar radius to $0.178 R_\odot$, consistent with the value that we adopt for the system. The dominant source of error in these relations is our choice of the stellar mass; improvements in the determination of the stellar mass will improve the determination of the stellar radius accordingly.

**Bolometric luminosity of the star.** We estimate the luminosity of LHS 1140 with the bolometric correction in $V$ and in $J$ from ref. 28. The absolute $M_J$ magnitude and bolometric correction[28] yield a luminosity of $0.00302 L_\odot$. Similarly, using the method of ref. 29, we find a luminosity of $0.002960 L_\odot$. We average these luminosities to estimate the stellar luminosity of LHS 1140 of $L_* = (0.002981 \pm 0.00021) L_\odot$. Using this luminosity estimate, we estimate the effective temperature of LHS 1140 to be $T_{eff} = 3{,}131 \pm 100$ K.

**Identification of the initial trigger.** The first observation of a transit of LHS 1140b by MEarth-South occurred in September 2014. However, we did not commence follow-up until October 2015. The long period of LHS 1140b makes it difficult to identify via traditional techniques such as phase-folded searches, which is how MEarth identified GJ 1214b (ref. 30) and quickly confirmed GJ 1132b (ref. 31). At these long periods, single-longitude observatories can expect to be able to observe only one or two transits per season during nighttime hours. This recovery rate is worsened by the likelihood of poor weather at the site during any of these events. Therefore, it is unlikely that MEarth-South would ever have been able to identify LHS 1140b from phase-folded searches of photometric data alone.

MEarth generates many triggers per night. The majority of these are false positives that can be explained through changes in the precipitable water vapour (which affects the differential brightness measurement between our 'red' target stars and our 'blue' reference stars). MEarth corrects for this by measuring the change in relative flux of all of the M dwarfs stars that are currently being observed, and constructing a 'common-mode' vector of the light-curve flux behaviour, which is common to all targets. However, the time-resolution of this common mode is 30 min, and more rapid changes in the precipitable water vapour at the observing site are not corrected and can potentially cause a MEarth trigger.

To sort through and discard these triggers, we trained a neural network[32] to classify MEarth data as either 'trigger' data or 'non-trigger' data, with the assumption that the majority of trigger data can serve as a proxy for data corrupted by systematic effects. We included MEarth triggers of GJ 1214 as non-trigger training events, because those observations were of a real transiting planet and not due to systematic effects. At the time we executed this method, we had not yet confirmed the existence of GJ 1132b, and therefore its triggers were omitted from this analysis entirely (because we also could not justify categorizing it as another false-positive). We included equal numbers of trigger observations and non-trigger MEarth observations in our training dataset. Non-trigger training data were selected such that, for each trigger that an individual target star contributed to the trigger training set, a random non-trigger data point from that same star was selected to be included in the non-trigger training set. Future work will include the triggers of GJ 1132b and LHS 1140b and could improve the performance of this analysis.

The nodes in our neural network are selected to be the weather and observatory state variables that we measure as part of routine MEarth observations. We selected the following state variables as the input nodes for our neural network: the atmospheric seeing, the ellipticity of the triggering image, the telescope pointing offset from the field's master image, the air mass, the zero-point offset (that is, extinction from clouds), the root-mean-square variation of the reference stars, the common mode, its derivative, and its offset from a 3-h mean. The MEarth common mode is a measure of the average change in magnitude of all of the M dwarf stars that MEarth is currently observing that night. This serves as a proxy for differential colour extinction associated with the change in water vapour concentration in the atmosphere. Including the derivative and the offset of this common mode can help to determine whether the data point is potentially affected by rapidly changing atmospheric conditions.

Our neural network is built to include one hidden layer of neurons, with the number of neurons in this hidden layer equal to the number of input parameters (described above). The neural network is allowed to train and converge via the back-propagation method. We did not investigate whether any of these input diagnostic parameters could be removed. We also did not investigate the effect of including different numbers of neurons in the hidden layer or including additional hidden layers.

After training this network, we selected all of the trigger events (approximately 10% of all MEarth-South triggers) that were misclassified as non-trigger data for further scrutiny. We visually inspected each of these triggers, and selected the trigger event from LHS 1140 to pursue.

**Photometric observations.** With the exception of the 1 September 2016 transit, all observations were gathered with the MEarth-South telescope array. Both 2014 trigger events were observed by MEarth-South telescope 1. MEarth-South telescopes 1 and 6 observed the June 2016 trigger event. This event was observed through clouds. After the June 2016 trigger event, a combined analysis of the 15 September 2014 trigger, the 19 June 2016 trigger and the radial-velocity observations collected to that date revealed three candidate periods for LHS 1140b. One of these candidate periods also back-predicted a partial transit observed by MEarth on 23 December 2014. We selected this ephemeris to predict future transit opportunities and we pursued four of these. We attempted to observe a full transit of LHS 1140b from Australia on 7 August 2016, but all sites were weathered out. We attempted to observe an additional transit from Australia and Hawaii on 1 September 2016. Five of the six sites were weathered out and we obtained a partial transit, including egress, from the Perth Exoplanet Survey Telescope (PEST). We succeeded in obtaining two full transit observations with MEarth-South on 25 September 2016 and 20 October 2016.

The 25 September 2016 and the 20 October 2016 full-transit events were observed by MEarth-

South telescopes 1, 2, 6 and 8. Exposure times were 23 s, yielding a cadence of 46 s for each of the MEarth-South telescopes for the high-cadence follow-up observations.

PEST is a home observatory with a 12-inch Meade LX200 SCT f/10 telescope with a SBIG ST-8XME CCD camera. The observatory is owned and operated by Thiam-Guan (TG) Tan. PEST is equipped with a *BVRI* filter wheel, a focal reducer yielding f/5, and an Optec TCF-Si focuser controlled by the observatory computer. PEST has a 31′ × 21′ field-of-view and a 1.2″ pixel scale. The observatory clock is synced on start-up to the atomic clock in Boulder, Colorado, and is re-synced every 3 h during an observing night. PEST observed an egress of LHS 1140b on 1 September 2016 in the I-band using 120-s exposures, and obtained a cadence of approximately 132 s.

We show each individual transit in Extended Data Fig. 1.

**Radial velocity observations.** We first gathered reconnaissance spectra of LHS 1140 with the TRES spectrograph at Fred Lawrence Whipple Observatory (FLWO). We measured a velocity shift between two spectra of 37 ± 34 m s$^{-1}$, which is consistent with 0 m s$^{-1}$, and began observations with the HARPS spectrograph on the La Silla 3.6-m telescope. We gathered 144 observations between 23 November 2015 and 13 December 2016, using an exposure time of 30 min and collecting two back-to-back exposures, although on three nights only one exposure was obtained. Each spectrum spans from 380 nm to 680 nm in wavelength. For each observation, we constructed a comparison template by co-adding all of the other spectra and measured the relative radial velocity as a shift required to minimize the $\chi^2$ of the difference between the spectrum and this template[33,34]. Telluric lines were masked using a template of lines made with a different, much larger dataset. We see no evidence for rotational broadening of the spectrum, consistent with the measured rotational period of LHS 1140. The median of the estimated internal error of the radial velocities is 4.1 m s$^{-1}$ per observation.

**Simple simultaneous photometric and radial velocity analysis.** We fit our light curves simultaneously with our radial velocities, letting the following parameters vary: orbital period $P$, time of mid-transit $T_0$, planet-to-star radius ratio $R_p/R_*$, radial-velocity semi-amplitude $K_b$, orbital inclination $i$, orbital eccentricity $e_b$, argument of periastron $\omega_b$, and separate baseline flux levels for each transit observed on each telescope. We simultaneously correct each light curve for a linear term in air mass to correct for differential colour extinction, and we also correct both full-transit observations with a linear trend in time, so that the out-of-transit baseline is equal on both sides of the transit.

Limb darkening was treated with a quadratic approximation, with coefficients initialized at values determined from model atmospheres[35]. During our analysis, we let these coefficients vary freely within 15% of these initial values. Owing to the similarity between the MEarth bandpass and the PEST bandpass, and to the limited amount of in-transit data for our PEST observation, we use the same limb-darkening parameters for all of our data. Our final limb-darkening parameters are $a = 0.219 \pm 0.025$ and $b = 0.415 \pm 0.083$.

The light-curve model is generated from JKTEBOP[36], modified to compute integrals analytically using the methods of ref. 37. We explore the parameter space with the emcee code[38], which is a Python implementation of the affine-invariant Markov chain Monte Carlo (MCMC) sampler, and take the 16th, 50th and 84th percentiles of the resultant parameters to obtain the 68% confidence interval for each of our model parameters.

For the radial velocities we adopt a range of functional forms to describe the radial-velocity signal caused by stellar activity, which we describe below.

We show the Lomb–Scargle[39,40] periodogram of the window function (the periodogram in which all radial-velocity measurements are set to 1) and of the measured HARPS radial velocities in Extended Data Fig. 2. Our window function contains substantial power at $P = 18$ days and at its harmonics, as well as near $P = 40$ days. Our measured radial velocities contain substantial power at the true orbital period ($P = 24.73712$ days) of LHS 1140b as well as power at higher periods (90 and 130 days). This excess power at high periods is due to radial-velocity variations induced by the stellar activity of LHS 1140. Stellar activity is known to produce systematic radial-velocity signal due to convective inhibition of starspots and to the flux imbalance between dark spots and bright faculae[41–43].

We assume a simple functional form for the radial-velocity variations induced by stellar rotation—a single sine wave. We initialize this sinusoid at the measured photometric rotational period of LHS 1140 (131 days) and an amplitude of 3 m s$^{-1}$, and allow these model parameters to vary freely. Our final best-fitting model has a radial-velocity rotational activity period of $129.5 \pm 4.5$ days, an amplitude of $4.26 \pm 0.60$ m s$^{-1}$ and an epoch of $2{,}457{,}641.1 \pm 3.6$ HJD. The period of this best-fitting sinusoid is consistent with the photometric rotational period. In Extended Data Fig. 3, we show the radial velocities of LHS 1140 phased to this period, along with our best fit.

When subtracting this activity-induced radial-velocity variation from our HARPS measurement, the Lomb–Scargle periodogram of our residual radial velocities contains reduced noise at long periods. Additionally, the highest remaining peak in this periodogram is the orbital period of LHS 1140b (see Extended Data Fig. 4a). After subtracting our best-fitting Keplerian orbit for LHS 1140b from these residual data, the Lomb–Scargle periodogram shows no more large peaks (Extended Data Fig. 4b). The residual peak near $P = 18$ days is due to the window function of our observations. We see no evidence for additional radial-velocity variations due to additional planets orbiting LHS 1140b, although additional observations and future work in which stellar velocity jitter due to activity is subtracted may improve the mass determination of LHS 1140b and potentially uncover additional signals from planets.

Radial velocity observations can also show substantial power at the harmonics of the stellar rotational period and impede planet detection[20]. To investigate the effect that this could have on our fit we performed an identical analysis, but also included sinusoidal terms with periods of $P_{\text{rotation}}/2$ and $P_{\text{rotation}}/3$, where $P_{\text{rotation}}$ is the stellar rotational period. The stellar rotational period is again initialized at the photometric rotational period and allowed to vary slightly. We find that, in this scenario, the best-fitting semi-amplitude for the radial-velocity signal of the planet has an amplitude of $6.0 \pm 0.7$ m s$^{-1}$, slightly higher than, but consistent with, the single-sinusoid model.

We also performed an identical fit, replacing the 131-day stellar rotational period and its harmonics with the 90-day peak seen in the radial-velocity periodogram and its harmonics. In this scenario, we find a semi-amplitude of $5.7 \pm 0.7$ m s$^{-1}$ for the radial-velocity variation induced by LHS 1140b, consistent with our other fits.

We note that for all of these analyses, we enforce a simple and strict functional form to describe the stellar contribution to the radial-velocity signal. In reality, this does not accurately model these effects, and so to obtain an uncertainty that more accurately reflects our uncertainties in the radial-velocity activity of the star, we turn to a more flexible model than we have presented here.

**Radial-velocity analysis with Gaussian process regression.** Features on the surface of the host star, such as dark spots, faculae and granulation, produce signals modulated by the stellar rotational period and its first and second harmonics[20]. These radial-velocity signals also evolve over time, and will thus produce quasi-periodic signals with varying amplitudes and phases. Therefore, the sinusoidal models above do not represent a perfect description of the stellar-induced radial-velocity variation. Theoretical[44,45] and observational[20,46] studies suggest that the surfaces of M dwarfs are covered with large numbers of small, low-contrast spots, which produce radial-velocity variations similar in amplitude to planetary-induced orbital motion, probably through inhibition of convection and flux imbalances. The exact shape of activity-induced radial-velocity signals, however, is still poorly understood. We see no evidence of correlations between radial velocities and the spectroscopic activity indicators (Ca ii H&K-derived S index and Hα emission) or between radial velocities and indicators of the asymmetry of the cross-correlation function (full-width at half-maximum and bisector span), although this is not unexpected in such a slowly rotating star[41]. Furthermore, the activity signals in radial velocities will not be perfectly correlated with photometric variations[20,47], which are sensitive to only the longitudinally asymmetric portion of the starspot distribution.

To account for activity-induced rotational modulation and evolution in the radial velocities we use Gaussian process regression[19], a technique that has been applied successfully in previous radial-velocity mass determinations of planets around active stars such as CoRoT-7[20,48], Kepler-78[49], Alpha Centauri B[50] and Kepler-21[51]. Gaussian process regression allows us to incorporate the uncertainty that arises from magnetic activity of the star in our determination of the planetary mass, while making minimal assumptions about the exact form of activity-induced radial-velocity signals.

Our Gaussian process has a quasi-periodic covariance kernel with timescales of recurrence $R$ and evolution $E$ that correspond to the rotational period and active-region lifetime of the star, respectively. We measure the rotational period from the light curve to be 131±10 days (which is also consistent with the $\log(R'_{HK})$-derived value of 127 ± 11 days, where $R'_{HK}$ is the chromospheric emission ratio, defined as the ratio of the fluxes from the Ca H and K lines to the broadband photometric flux in the R band). The periodogram of the radial velocities (Extended Data Fig. 2b) displays a peak not only around 130 days, but also at $P$ = 90 days; however, this peak is present in the window function (Extended Data Fig. 2a), which means that it is probably a fake signal that arises from our observing strategy rather than a true stellar signal. To prevent hyperparameter $R$ from settling at this 90-day value, we impose a Gaussian prior with a slightly narrower width of 5 days instead of 10 days. The light curve is too short to give a precise estimate of the average lifetime of the active regions on the stellar surface, but we can infer that it is longer than a few rotational cycles; we adopt a value of three times the stellar rotational period (393 days). We constrain hyperparameter $E$ with a Gaussian prior centred at this value, with a width of 30 days. We find that choosing a different value (of the same order of magnitude) does not greatly affect our mass determination and its associated uncertainty.

Another parameter, $S$, determines the maximum number of peaks that the activity-induced radial-velocity signal is allowed to have within each recurrence timescale. Radial-velocity curves typically show two or three peaks per rotation (limb effects and degeneracies wash out small-structure signatures), which can be reproduced with $S$=0.5. We are therefore justified in adopting a strong Gaussian prior for this parameter (0.5 ± 0.05).

The amplitude $A$ of the Gaussian process is treated as a free parameter, drawn from a uniform and positive prior distribution. The orbit of the planet is modeled as a Keplerian orbit with free eccentricity, with orbital period and time of transit constrained by Gaussian priors based on our transit analysis. The radial-velocity semi-amplitude $K_b$, is drawn from a uniform, positive prior distribution. We account for the presence of additional uncorrelated, Gaussian noise with a term added in quadrature to the diagonal elements of the covariance matrix. These diagonal elements correspond to the variance of each observation, which are routinely taken as the formal radial-velocity uncertainties. It is drawn from a Jeffreys prior distribution with a lower bound equal to the internal instrumental precision of HARPS (0.6 m s$^{-1}$).

The best-fitting parameters and their associated uncertainties are determined through an MCMC procedure akin to that described in ref. 51. Our best-fitting radial-velocity model is shown in Extended Data Fig. 5 and Extended Data Fig. 6 shows correlation plots for all of the parameters of our radial-velocity model; no substantial correlations are present and all of the parameter chains achieve good convergence.

We measure a radial-velocity semi-amplitude of $K_b = 5.34 \pm 1.1$ m s$^{-1}$ for LHS 1140b, which corresponds to a mass of $M_p = (6.65 \pm 1.82) M_\oplus$. We place an upper limit on the eccentricity of $e_b < 0.29$ with 90% confidence. We adopt these values as our final values for the physical parameters of the LHS 1140 system.

**Rejection of false-positive scenarios.** Owing to the high proper motion of LHS 1140, we are able to inspect the POSS-I E photographic plates for luminous background objects located at the current position of LHS 1140. These plates have a magnitude limit of 19.5 in a band pass roughly similar to the R band, approximately 9 magnitudes fainter than LHS 1140. We see no source at the current position of LHS 1140, and therefore conclude that there is no background source contaminating our transit measurements or producing a false-positive signal.

To investigate the possibility of a luminous or massive bound companion in the LHS 1140 system, we inspected the high-resolution spectra taken with HARPS for signs of additional radial-velocity drift or of additional spectral features. We see no evidence for a radial-velocity drift over these observations or any additional spectral features.

On the basis of the lack of a reported perturbation on the parallax[14] and the ten-year span of data, we can rule out any signal due to a brown dwarf companion. A bound L0 brown dwarf at 5 au would be approximately 6 magnitudes fainter in $V$, but would induce a perturbation of 175 mas, twice the parallax amplitude of LHS 1140. Given these strict astrometric limits on bound companions, we reject false-positive scenarios and substantial contamination by a third light source in our transit measurements.

**The eccentricity of LHS 1140b.** We have ruled out eccentricities larger than $e_b = 0.29$ for LHS 1140b, and our observations are consistent with a circular orbit. Unlike the other small planets that orbit mid-to-late M dwarfs and have well-con- strained eccentricities (GJ 1214[30] and GJ 1132[31]), tidal circularization is inefficient at the orbital distance of LHS 1140b, and its circular orbit is probably natal. This means that LHS 1140b either formed in its present location or, if it migrated, this migration must have proceeded smoothly in the protoplanetary disk. More violent migration mechanisms, such as planet–planet scattering, would have produced a system with large orbital eccentricity. Even though we constrain the eccentricity of LHS 1140b to be low, future determination of its precise value is important for discussions of its climate and potential habitability.

**Metallicity of the star.** We gathered a near-infrared spectrum of LHS 1140 with the IRTF/SPeX instrument[52]. We used an observing time of 60 s per exposure, with 8 total exposures (478 seconds of total integration), with the 0.3″ × 15″ slit and a resolving power of $R = 2,000$. We used HD 3604 as our AOV standard star. We compared this spectrum by eye to the IRTF Spectral Library[53,54] and classified LHS 1140 as an M4.5V-type star (Extended Data Fig. 7).

We use near-infrared spectral features that have been shown to be sensitive indicators of stellar metallicity[55,56] to estimate the metallicity of LHS 1140. For features located in the H band, we find a value of [Fe/H] = −0.04 ± 0.10 and [M/H] = −0.24 ± 0.09; for features in the $K_s$ band, we find [Fe/H] = −0.30 ± 0.08 and [M/H] = −0.25 ± 0.08. An estimate of the MEarth-band magnitude of LHS 1140, combined with our photometric metallicity relation[57], yields an estimate of [Fe/H] = −0.22 ± 0.1. All of these relations suffer from a relative lack of low-metallicity calibrator systems, and the uncertainty is probably higher than the internal errors reported for each index. However, with the exception of [Fe/H] indicators in the H band, we consistently find that LHS 1140 is a metal-poor star compared to the Sun, and we adopt a metallicity of [Fe/H] = −0.24 ± 0.10, where this error is the standard deviation of the individual estimates and does not account for any systematic errors internal to the calibrations themselves. We estimate this systematic error to be of the same magnitude as the scatter in the relation itself.

**Rotational period and age of the star.** MEarth data are calibrated each night using observations of standard star fields. Analysis of these data[58,59], calibrating each night and correcting for the effects of camera changes and the slow loss of sensitivity from dust settling on the instruments, has allowed us to measure the photometric rotational periods for many of our targets. We measure a photometric modulation, which we interpret as the rotational period, of 131 days for LHS 1140b (Extended Data Fig. 8). This rotational period is consistent throughout the MEarth-South observational seasons and is consistent between telescopes. We estimate a rotational period from $\log(R'_{HK})$ (ref. 60) of 127 ± 11 days, consistent with our photometric measurement. We also find that the amplitude of the photometric modulation due to rotation has increased between the 2014–2015 observing season and the 2016 observing season. Because this signal is determined by the longitudinal asymmetric distribution of starspots, it is difficult to tell whether this is an indication of a long-term trend, part of a stellar cycle, or simply due to a slight increase in starspot asymmetry over the MEarth-South observational baseline. M dwarfs spin more slowly as they age, with less-massive stars reaching longer rotational periods at older ages than more massive stars[61]. Barnard's star has a rotational period of 130 days (ref. 62), a mass of $0.16 M_\odot$ and is estimated to be 7–13 Gyr old[62]. This suggests that LHS 1140 is likely to be comparable in age to Barnard's star.

LHS 1140 has a heliocentric (U, V, W) velocity of (3.64 ± 0.42, −40.77 ± 1.35, 5.40 ± 0.27) km s$^{-1}$. The Sun's motion relative to the local standard of rest has been measured to be ($11.1^{+0.69}_{-0.75}$, 12.24 ± 0.47, 7.25 ± 0.37) km s$^{-1}$ (ref. 64). Correcting to the local standard of rest, we find that LHS 1140 has a (U, V, W) velocity of ($-7.46^{+0.80}_{-0.86}$, −53.01 ± 1.43, −1.85 ± 0.46) km s$^{-1}$ and is consistent with a kinematically older stellar population[65], although this is not a tight constraint.

Furthermore, we detect no Hα emission in our spectra. M4.5V dwarfs tend to have active Hα emission for the first 5 Gyr of their lives[66,67]. On the basis of gyrochronology and asteroseismology relations[68], the age of Alpha Centauri B is estimated to be approximately 6–7 Gyr. The rotational period of Proxima Centauri is approximately 86 days (ref. 69), suggesting that LHS 1140 may be older than Proxima Centauri. Taken together, we suggest that LHS 1140 is likely to be greater than 5 Gyr in age.

**Code availability.** The EMCEE code is available as a Python install and is publicly available on GitHub (https://github.com/dfm/emcee). JKTEBOP and the modifications performed to it are also publicly available on GitHub (https://github.com/ mdwarfgeek/eb). These codes were used to produce the light curve and radial- velocity models used in the analysis of our data.

The code used to determine the planetary orbit and mass, with Gaussian process regression to account for the magnetic activity variations of the host star, is not yet publically available, but we are currently working to make it so.

**Data availability.** All data used in this work are provided as Supplementary Data, and are available on the MEarth project webpage (https://www.cfa.harvard.edu/ MEarth/Welcome.html) and via the online repository FigShare (https://figshare. com/s/9e2b29d4f7a8043ca071 for MEarth and PEST photometry; https://figshare. com/s/49625e95aabf9e1f2ae6 for HARPS radial-velocity data).

**Additional References**

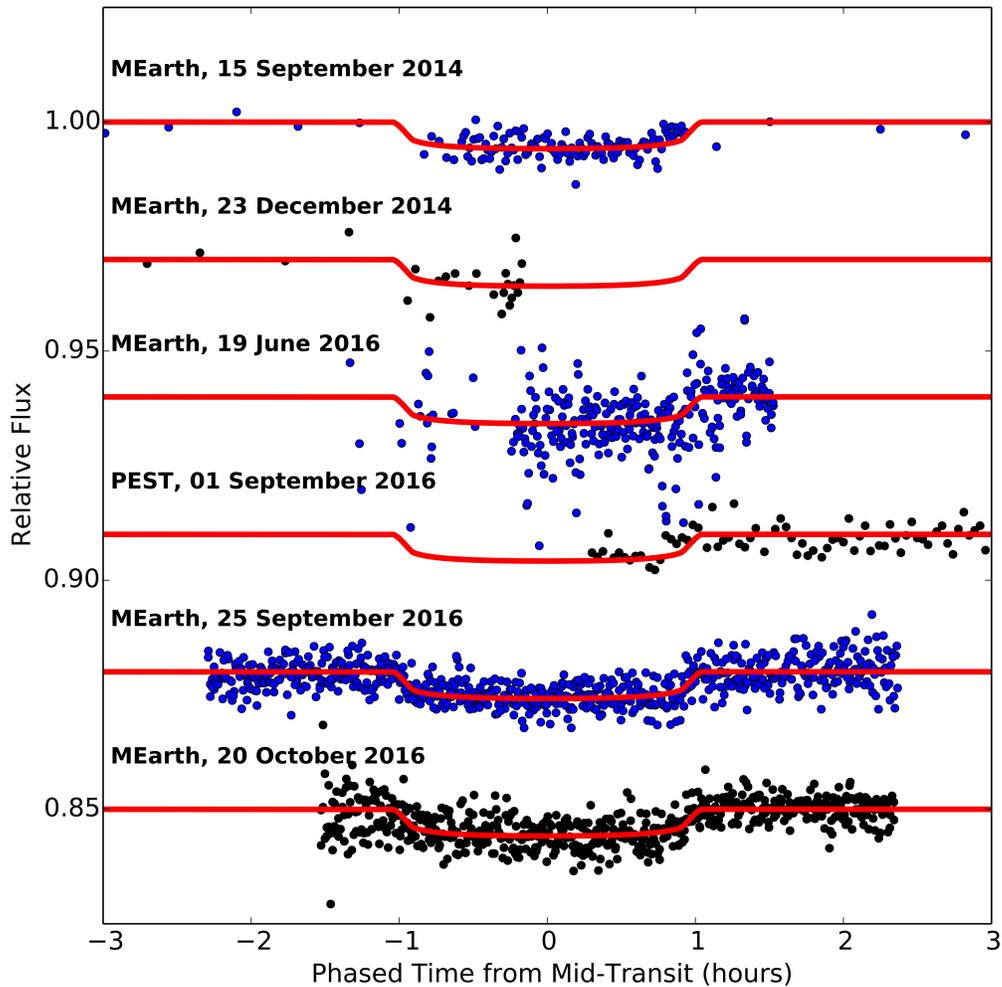

***Extended Data Figure 1 | Individual transit light curves observed with MEarth and PEST.*** *The top three light curves are MEarth-South trigger observations whereas the bottom three are targeted transit observations. Light curves are offset and shown in alternating colours for clarity. We have corrected for the effects of air mass for all light curves and have fitted a linear trend in time to both full-transit observations (the bottom two light curves).*

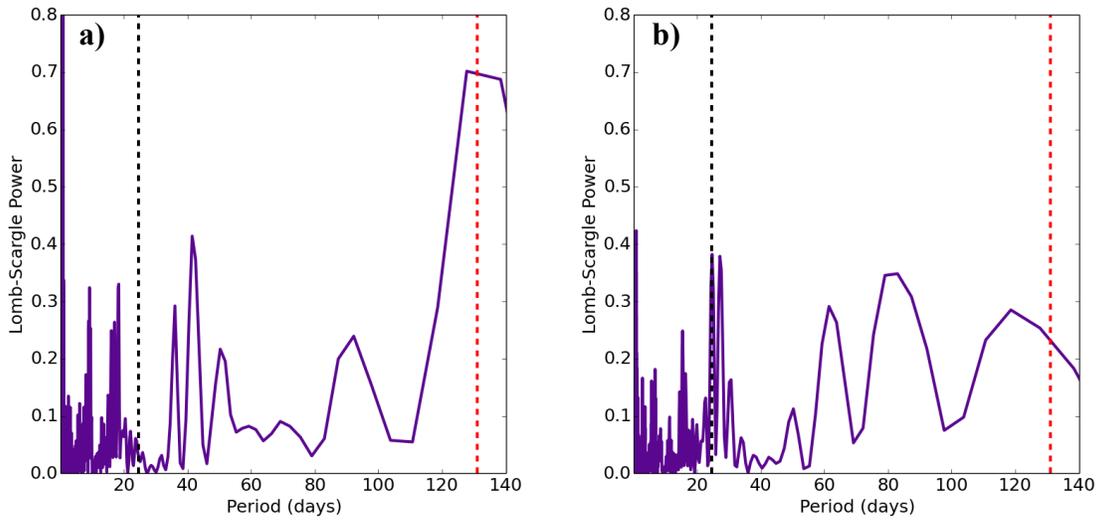

*Extended Data Figure 2 | Lomb–Scargle periodograms of HARPS radial-velocity measurements. a*, Periodogram of our window function for our radial-velocity observations. We see large peaks at approximately 18 days and its harmonics. We also see large peaks at around 40 daysand at the stellar rotational period of 131 days (red dashed line). The 24.73712-day orbital period of LHS 1140b is indicated by the black dashed line. *b*, Periodogram for our HARPS radial-velocity observations. We see substantial power near the 24.73712-day orbital period of LHS 1140b (black dashed line), as well as broad power at longer periods associated with stellar rotation (131 days, red dashed line) and at shorter periods associated with the window function of our observations.

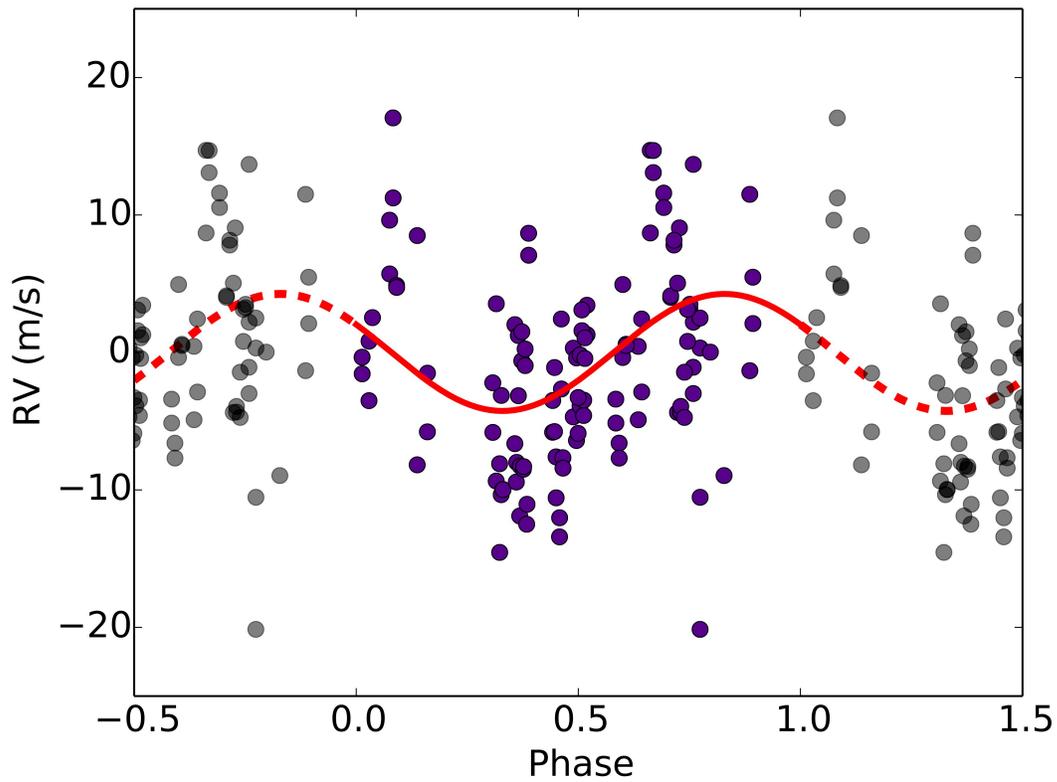

**Extended Data Figure 3 | Radial-velocity (RV) observations of LHS 1140 phased to our best-fitting sinusoid for the stellar activity signal.** The signal from LHS 1140b has not been removed. Grey data points are copies of the purple data points and the red line represents our best-fitting model. We find a radial velocity amplitude of 4.26 ± 0.60 m s$^{-1}$ due to stellar activity coupled with rotation, comparable to the radial- velocity amplitude of LHS 1140b. RV = 0 corresponds to the radial velocity of the host star.

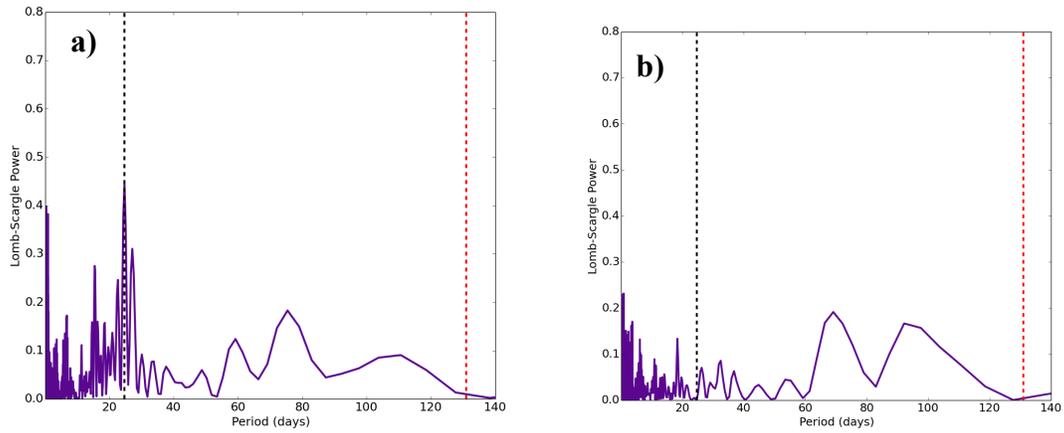

***Extended Data Figure 4 | Lomb–Scargle periodograms of HARPS radial velocity measurements. a***, *Periodogram of our residual radial velocities after subtracting our best-fitting sinusoid for the stellar activity signal. The highest peak in this dataset is located at the orbital period of LHS 1140b (black dashed line) and the broad power located at long periods has been suppressed.* ***b***, *Periodogram of our residual radial velocities after subtracting our best-fitting model that includes stellar radial-velocity variation as well as the orbit of LHS 1140b. We see no additional substantial peaks in our radial velocities, with the small peak located near P = 18 days due to the window function of our observations.*

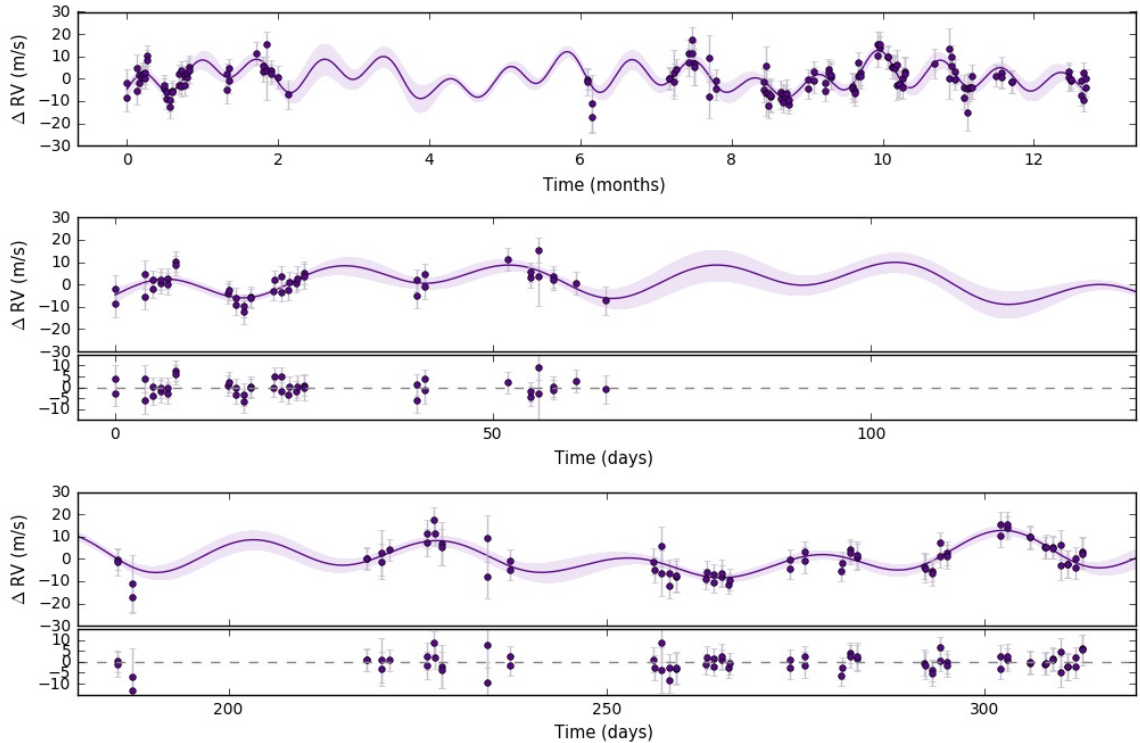

*Extended Data Figure 5 | HARPS radial-velocity (ΔRV) measurements fitted with a Gaussian process model. Our radial-velocity observations (points; error bars are 1σ) and best-fitting Gaussian-process-based model (line with shaded 1σ error regions). The high cadence and adequate observational strategy allows us to identify the orbital signature of the planet by eye, and the Gaussian-process-based model naturally incorporates the uncertainty that arises from the activity-driven signals of the host star in our final determination of planetary mass. The residuals obtained after subtracting the model from the data are shown for the bottom two panels. ΔRV = 0 corresponds to the radial velocity of the host star.*

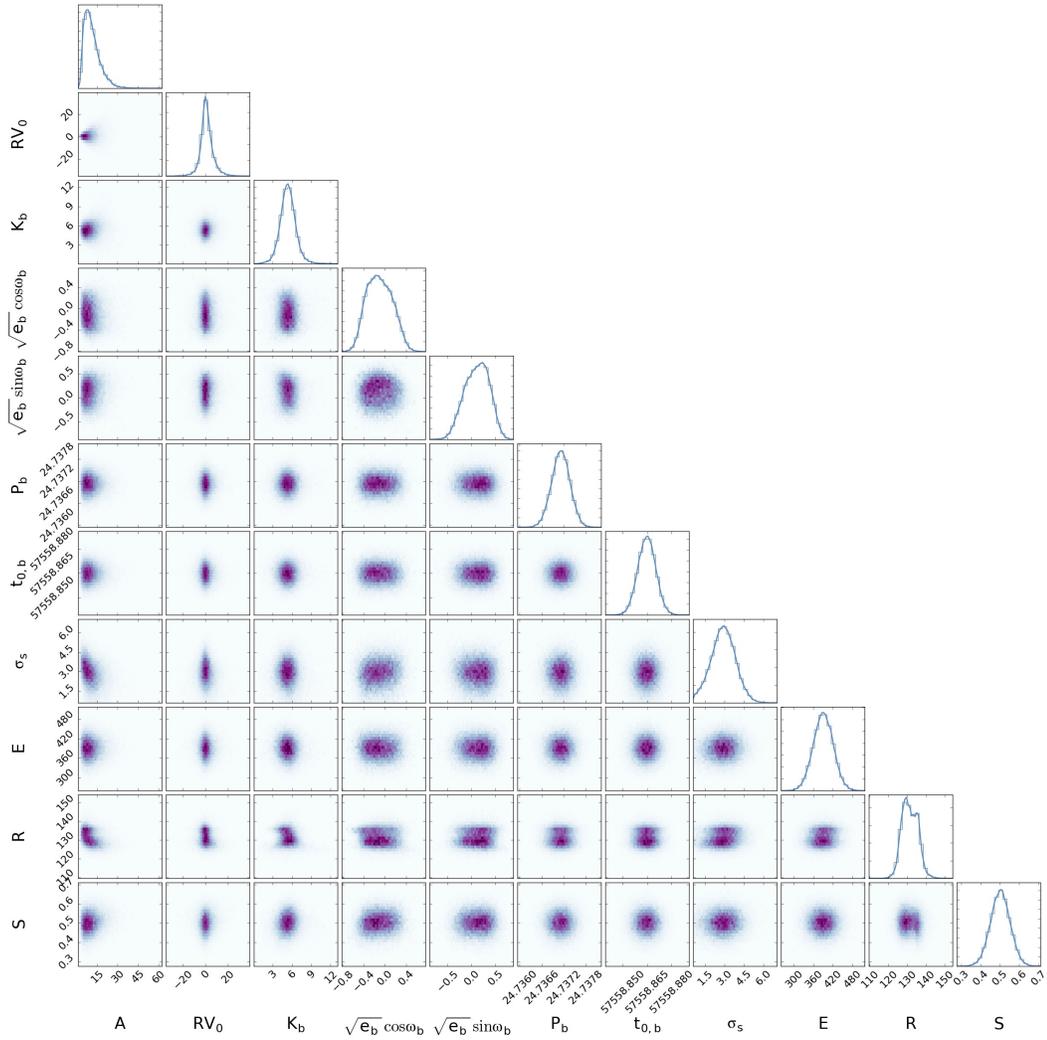

*Extended Data Figure 6 | Marginalized posterior distributions of the radial-velocity model parameters.* The solid lines over-plotted on the histograms are kernel density estimations of the marginal distributions. These smooth, Gaussian-shaped posterior distributions indicate the good convergence of the MCMC chain. Here we show the parameters of the Gaussian process (A, E, R and S), the orbital period of LHS 1140b ($P_b$), its radial velocity (relative to the host star; $RV_0$), radial-velocity semi-amplitude ($K_b$) and time of mid-transit ($t_{0,b}$), the parameters $e_b \sin(\omega_b)$ and $e_b \cos(\omega_b)$, where $e_b$ is the eccentricity of LHS 1140b and $\omega_b$ is its argument of periastron, and a white-noise error term ($\sigma_s$; in units of metres per second) that is added in quadrature to each HARPS radial-velocity data point. The top left is histogram is for A (the amplitude of the Gaussian process).

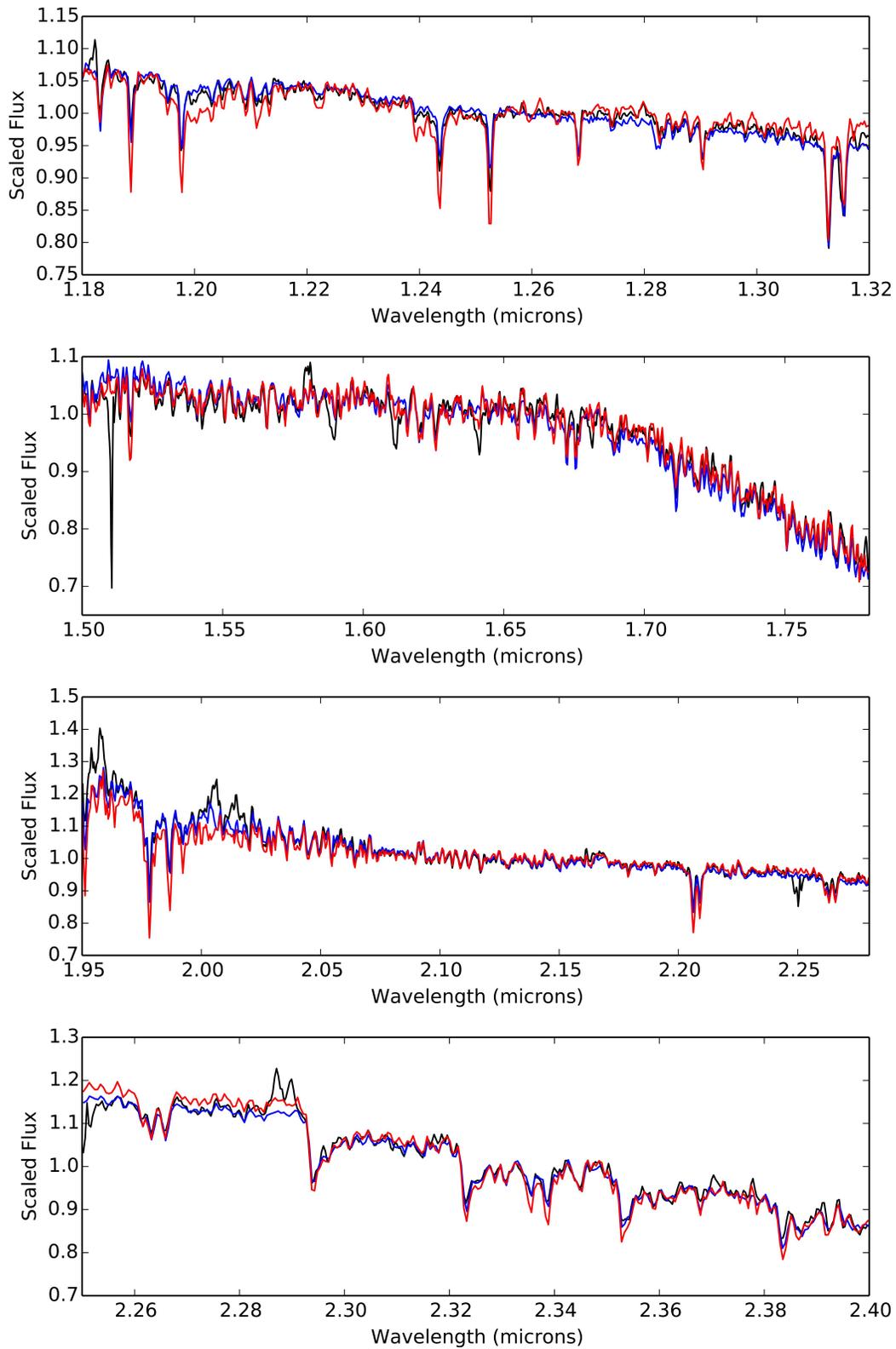

***Extended Data Figure 7 | Near-infrared spectrum of LHS 1140.*** *IRTF/SPeX spectrum of LHS 1140 (black), with M4V (blue) and M5V (red) spectra over-plotted. We classify LHS 1140 as an M4.5V star on the basis on its near-infrared spectrum.*

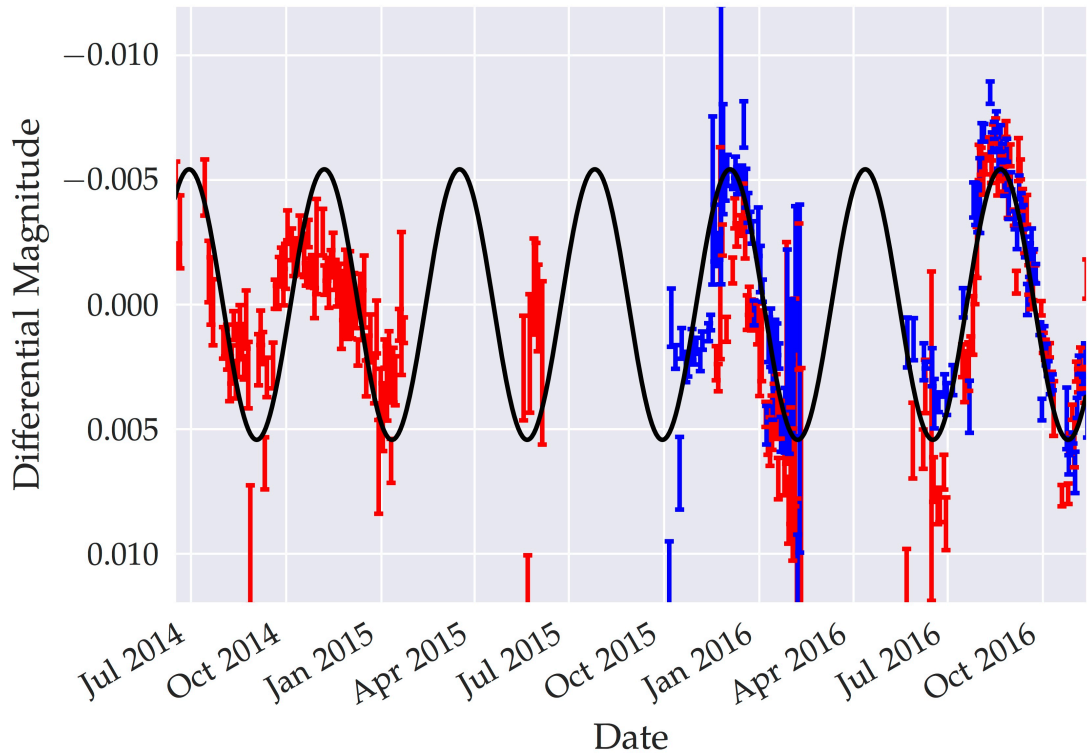

**Extended Data Figure 8 | Photometric modulation from the rotation of LHS 1140.** *The plot shows MEarth-South photometry of LHS 1140. We find that LHS 1140 has photometric modulation due to stellar rotation and the asymmetric distribution of starspots, with a period of 131 days. Data from the two telescopes monitoring LHS 1140 are coloured in red and blue, and a sinusoid with a 131-day period is over-plotted in black. We find that the amplitude of variation increased between 2014 and 2016. Error bars, 1σ.*